\newcommand{\nphysa}{Nucl. Phys. \textbf{A}}

\newcommand{\ffg}{\mathrm{FFG}}

\newcommand{\cl}{\mathrm{cl}}

\newcommand{\WS}{\mathrm{WS}}

\newcommand{\oem}{\mathrm{oem}}

\newcommand{\sat}{\mathrm{sat}}

\RequirePackage{fix-cm}
\documentclass[smallextended]{svjour3}       
\smartqed  
\usepackage{graphicx}
\begin{document}

\title{Properties of neutron star crust with improved nuclear physics 
}
\subtitle{Impact of chiral EFT interactions and experimental nuclear masses}


\author{Guilherme Grams         \and
        J\'er\^ome Margueron   \and
        Rahul Somasundaram     \and
        Sanjay Reddy
}


\institute{G. Grams, J. Margueron, R. Somasundaram \at
              Univ Lyon, Univ Claude Bernard Lyon 1, CNRS/IN2P3, IP2I Lyon, UMR 5822, F-69622, Villeurbanne, France
            \and
            S. Reddy \at
Institute for Nuclear Theory, University of Washington, Seattle, WA 98195-1550, USA\\ JINA-CEE, Michigan State University, East Lansing, MI, 48823, USA            
}

\date{Received: date / Accepted: date}

\maketitle

\begin{abstract}
A compressible liquid-drop approach adjusted to uniform matter many-body calculations based on chiral EFT interactions and to the experimental nuclear masses is used to investigate the neutron star crust properties.
Eight chiral EFT hamiltonians and a representative phenomenological force (SLy4) are confronted.
We show that some properties of the crust, e.g. clusters mass, charge, and asymmetry, are mostly determined by symmetric matter properties close to saturation density and are therefore mainly constrained by experimental nuclear masses, while other properties, e.g., energy per particle, pressure, sound speed, are mostly influenced by low-density predictions in neutron matter, where chiral EFT and phenomenological forces substantially differ.
\end{abstract}

\section{Introduction}
\label{sec:intro}

Neutron stars (NS) are excelent laboratories to test nuclear physics models under extreme conditions of density and isospin asymmetry~\cite{Rezzolla2018}. They are mainly divided into two regions: the core which contains about 99\% of the mass and the crust. In this study, we will focus on the properties of the crust since it coincides with the density region where chiral EFT approach is well defined. The upper bound in density of the crust is connected to the saturation energy density of nuclear matter, $\rho_\sat\approx 2.6\times 10^{14}$~g~cm$^{-3}$.
The outer crust contains a mixture of electrons and nuclear clusters, complemented with a neutron fluid in the inner crust. This neutron fluid is the unique physical realization of neutron matter (NM), although it is not a pure uniform matter phase.
From a nuclear theory view point, NM presents several advantages which makes it easier to model and accurate to predict. One may then wonder how the properties of NM, which are predicted by several modelings, impact those of the crust, which can be related to observations.

Several important and conceptual steps have recently been performed in the prediction of NM from microscopic approaches based on the bare nuclear interaction, i.e. the interaction between nucleons determined by free nucleon scattering up to a few hundreds of MeV. The main one was carried by the advance of nuclear interactions capturing low energy QCD symmetries in an effective way (chiral effective field theory, $\chi_{EFT}$) and called chiral NN interactions. This approach also generates many-body interactions consistently with the order considered for the two-body interaction. They have been applied to predict homogeneous nuclear matter properties up to the break-down density, i.e. the maximal density at which $\chi_{EFT}$ approach can be used, from various many-body approaches \cite{Carbone2013,Hebeler2015,Drischler2019,Lynn2019,Rios2020}. At sufficiently low densities, NM is well understood because three-body interactions are small, and the two-body neutron-neutron interaction is strongly constrained by neutron-neutron scattering phase shifts~\cite{Heiselberg2000,Carlson2003}.
The precise location of the break-down density is still a matter of debate, and it is presently expected to be at around saturation density ($n_\sat\approx 0.155$~fm$^{-3}$) or slightly above~\cite{Epelbaum2009,Tews2018,Drischler2020}. Note that $\rho_\sat$ and $n_\sat$ represent the same quantity in two different units.

Some properties of the inner crust are mainly constrained by those of NM, which implies that the inner crust can be viewed as a uniform distribution of neutrons and electrons modified by a few impurities (nuclear clusters), while some other properties reflect the cluster properties, which makes the inner crust viewed as a continuation of the nuclear chart, where isolated nuclei becomes dilute nuclei surrounded by a shallow neutron fluid. The scope of this study is to analyze which properties of the inner crust are mostly impacted by the NM modeling, and can thus be considered as a way to probe the predictions from $\chi_{EFT}$, and which ones mainly reflect the properties of isolated nuclei, for which accurate experimental data exist.

To do so, one needs a modeling for the NS crust accurate enough in its prediction, but which allows as well a clear understanding of the different contributions to the equation of state (EoS). In their seminal paper in 1971, Bethe, Baym, and Pethick~\cite{Baym1971} have introduced the first version of the compressible liquid-drop model (CLDM) with finite-size (FS) contribution containing Coulomb and surface terms. Latter on Douchin and Haensel have developed further this model by considering in 2001 the curvature term contribution to the FS effects~\cite{DouchinHaensel2001}, determined from many-body methods~\cite{DouchinHaensel2000}.
In our study, we adopt the CLDM approach, as in Ref.~\cite{Carreau2019a}, where the bulk term is fixed by the nuclear meta-model~\cite{Margueron2018a} and the FS term contains Coulomb, surface and curvature contributions as described by the FS4 approximation presented in Ref.~\cite{Grams2021}.
The FS terms may however still be fine tuned on experimental nuclear masses, as suggested by Steiner~\cite{Steiner2008} since they are derived in the CLDM by neglecting the smoothing effect of the surface.
The meta-model is adjusted to reproduce the many-body perturbation theory predictions in uniform matter from $\chi_{EFT}$ NN and 3N interactions, as first suggested by Tews~\cite{Tews2016}.
With such a CLDM constrained by uniform matter calculations based on several $\chi_{EFT}$ interactions on one side, and by the experimental nuclear masses on the other side, the modeling of the crust properties is qualitatively well under control and the connection between the different ingredients in the model and the prediction is rather clear. 

In the following, we first confront uniform matter calculations based on $\chi_{EFT}$ interactions to the ones based on a representative phenomenological force such as SLy4~\cite{Chabanat1997}. We then show that the fits to experimental nuclear masses disfavor some of the $\chi_{EFT}$ Hamiltonians. Finally, we compute NS crust properties, discuss their connection to NM predictions and to the experimental nuclear masses, and confront our best models to well-known EoS.

\section{Uniform matter predictions based on $\chi_{EFT}$ interactions}
\label{sec:um}

\begin{figure*}[t]
\centering
\includegraphics[scale=0.29]{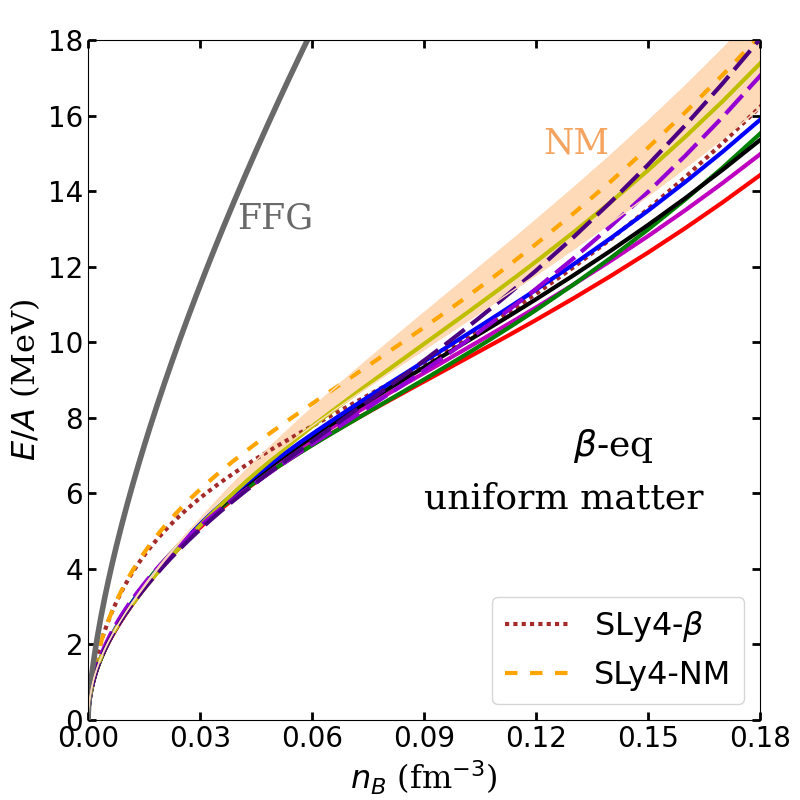}
\includegraphics[scale=0.29]{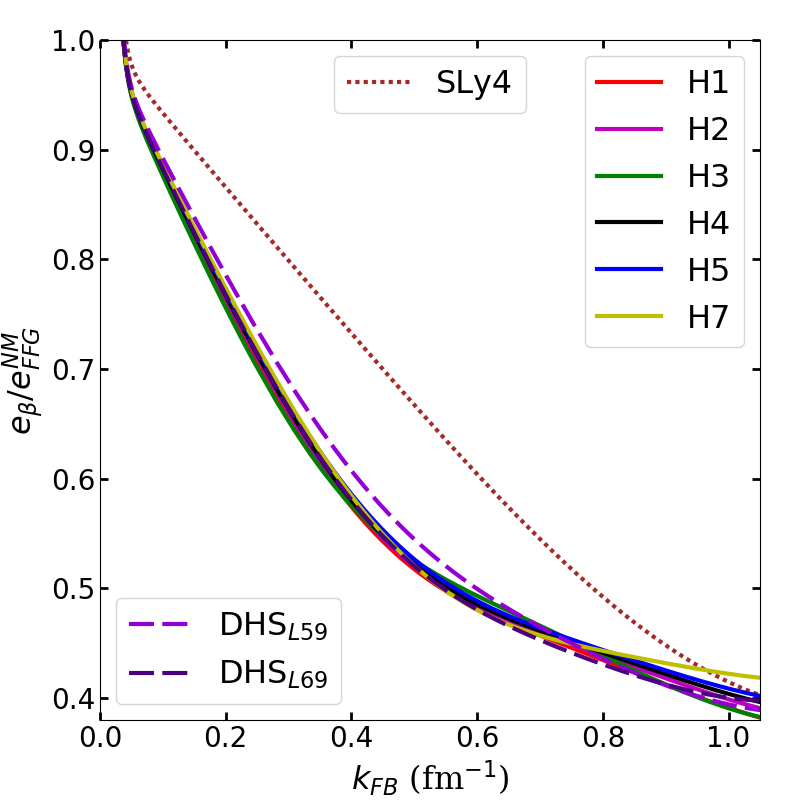}
\caption{Left: $\beta$-equilibrium energy per particle as function of the baryon density for the  $\chi_{EFT}$ Hamiltonians (the six hamiltonians H1-H7~\cite{Drischler2016}, except H6, and the two recent ones DHS$_{L59}$ and DHS$_{L69}$~\cite{Drischler2021}) and the SLy4 force~\cite{Chabanat1997} (NM: dashed line, $\beta$-equilibrium: dotted line), pale orange band corresponds to the NM energy per particle prediction for H1-H7. Right: $\beta$-equilibrium energy per particle normalized to the NM Free Fermi Gas energy $E_\ffg^\mathrm{NM}$ function of the Fermi momentum $k_{FB}$.}
\label{fig:enerBetaUnif}
\end{figure*}

In a recent paper~\cite{Grams2021}, we have presented in detail the CLDM that we use in the present study and detailed how the bulk properties are connected to uniform matter properties determined from $\chi_{EFT}$ interactions, by using the meta-model approach~\cite{Margueron2018a}. 
The $\chi_{EFT}$ predictions that we are analyzing are the many-body perturbation predictions based on the H1-H7 hamiltonians~\cite{Drischler2016}, except H6 for which the $^3$H binding energy is not well fit~\cite{Drischler2016}, and the two DHS (DHS$_{L59}$ and DHS$_{L69}$) recent predictions~\cite{Drischler2021}.

The $\beta$-equilibrium properties in uniform matter, derived from the $\chi_{EFT}$ approach as well as by the SLy4 force are shown in Fig.~\ref{fig:enerBetaUnif}.
The energy per particle for H1-H7 (except H6), for DHS Hamiltonians and for SLy4 are shown first as function of the density (left panel) and then the energy per particle normalized to the FFG energy $E_\ffg^\mathrm{NM}=3\hbar^2k_{Fn}^2/(10m)$, with the neutron Fermi momentum $k_{Fn}=(3\pi^2 n_B)^{1/3}$, as function of the isoscalar Fermi momentum $k_{FB}=(3\pi^2 n_B/2)^{1/3}$ (right panel).
As expected, at low densities the predictions of all $\chi_{EFT}$ converge to a narrow band, while SLy4 predicts slightly larger values for the energy per particle. The differences between SLy4 and chiral EFT predictions at low density are better observed on the right panel where the energy per particle is normalized to the Free Fermi Gas energy. 
SLy4 predicts up to 30\% larger energy per particle compared to the $\chi_{EFT}$ group for $0.1<k_{FB}<0.9$~fm$^{-1}$, the largest differences being observed for $k_{FB}\simeq 0.4$~fm$^{-1}$.
In comparison to this difference, the $\chi_{EFT}$  predictions are much closer for $k_{FB}<0.9$~fm$^{-1}$. The dispersion among the $\chi_{EFT}$ predictions is less or about a few percent for $k_{FB} < 0.9$~fm$^{-1}$ and above the dispersion is larger than the difference between SLy4 and chiral EFT.

The question which then arises is whether this systematic difference between the group of $\chi_{EFT}$ predictions and SLy4, which is a typical prediction for phenomenological models, impacts the NS crust properties. And if yes, for which specific properties?

\begin{figure*}[t]
\centering
\includegraphics[scale=0.29]{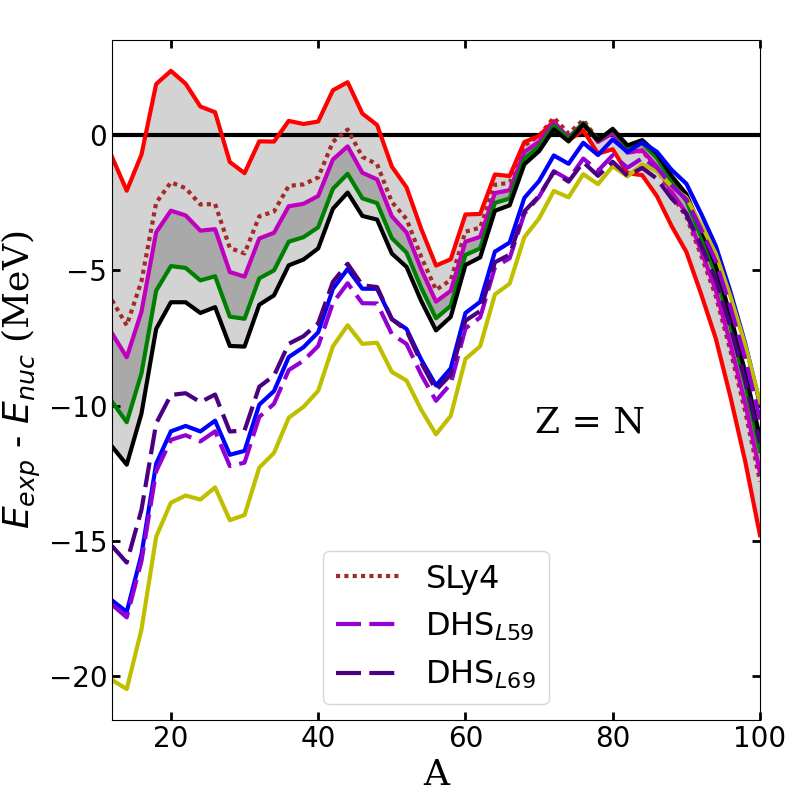}
\includegraphics[scale=0.29]{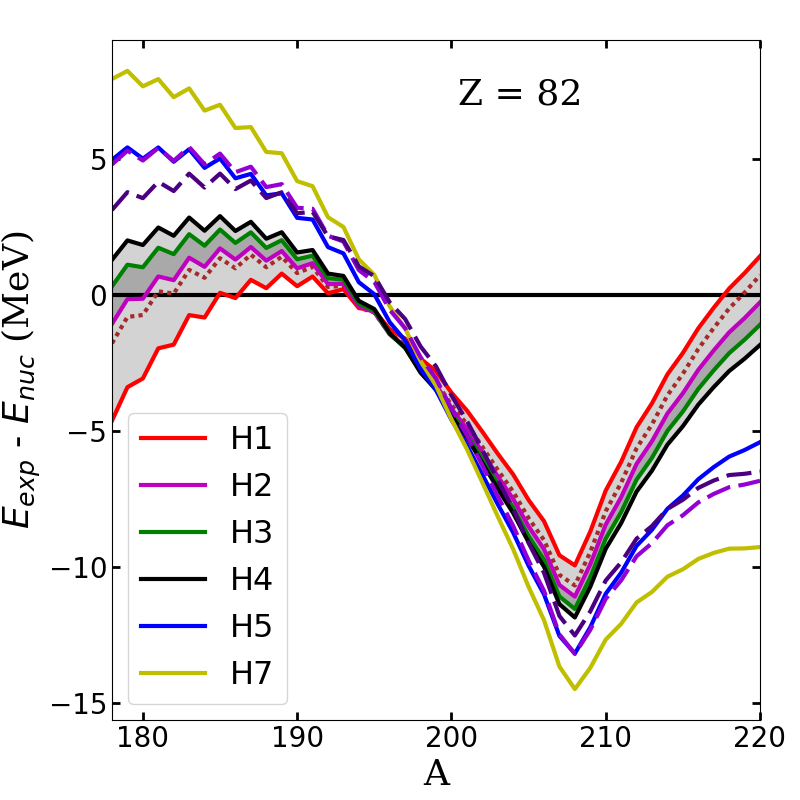}
\caption{Left: Residuals for $N=Z$ nuclei for SLy4 and $\chi_{EFT}$ Hamiltonians. Right: Residuals for Pb isotopes ($Z=82$). Light grey band shows the selected Hamiltonians H1-H4, dark grey band shows the two best Hamiltonian (H2 and H3) when confronted to the experimental nuclear masses.
For clarity, the experimental data are corrected by the odd-even mass staggering as: $E_{\cl}^{\oem}(A,I)  = a_{\oem} A^{-1/2} \delta_{np}$, where $a_{\oem}= -12$~MeV and $\delta_{np}=1$ if $N$ and $Z$ odd, $0$ if $N$ or $Z$ odd, and $-1$ if $N$ and $Z$ even~\cite{BohrMottelson1969}. }
\label{fig:nucl:res}
\end{figure*}

\section{Application of the CLDM to finite nuclei}
\label{sec:finitenuclei}

Before addressing the links between $\chi_{EFT}$ predictions in NM and NS crust properties, we briefly illustrate the role played by finite nuclei to constrain the CLDM.

We confront in Fig.~\ref{fig:nucl:res} the CLDM prediction for finite nuclei energies with the experimental masses~\cite{AME2016} for symmetric nuclei (left panel) and for neutron rich Pb isotopes fixing $Z=82$ (right panel). The ability of the models H1-H4 to reproduce low- and high-A is quite uniform, while the models H5, H7 as well as DHS$_{L59}$ and DHS$_{L69}$ depart from the trend given by the best hamiltonians, at low-$A$ and for asymmetric nuclei (neutron rich or neutron poor). The confrontation with the experimental nuclear masses allows us to select H1-H4 hamiltonians (with light-gray band) as being our optimal models, see Ref.~\cite{Grams2021} for more details, with H2-H3 (dark-gray band) being the best. Note that the phenomenological SLy4 force reproduces experimental nuclear masses with an accuracy equivalent to our best models.

\begin{figure*}[t]
\centering
\includegraphics[scale=0.35]{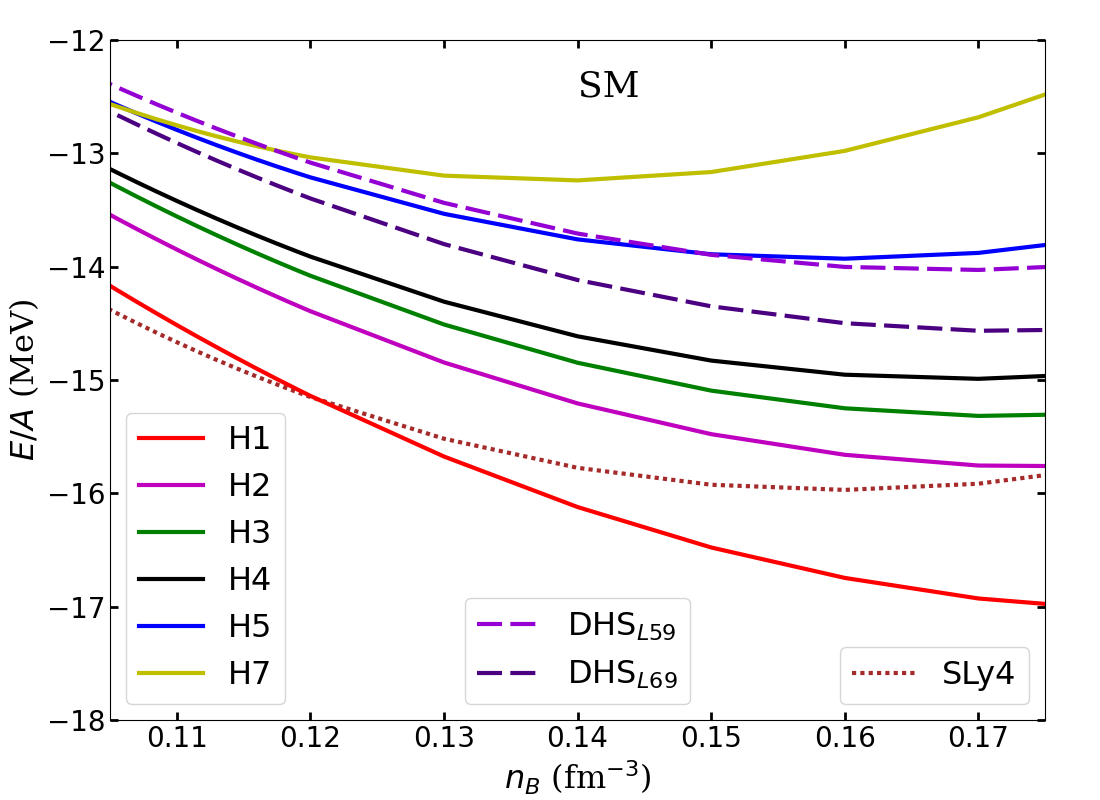}
\caption{Energy per particle as function of the baryon density around saturation density $n_\sat\approx 0.16$~fm$^{-3}$ in symmetric matter. There is correlation between the energy per particle in SM at $n_\sat$ and the residuals shown in Fig.~\ref{fig:nucl:res}.}
\label{fig:enerSM}
\end{figure*}

It is interesting to connect these results with the energy per particle predicted by these models in symmetric matter (SM) shown in Fig.~\ref{fig:enerSM}.
Even if the contribution from the FS terms (Coulomb and surface) are absent in uniform matter, one could observe a clear correlation between the CLDM predictions for $N=Z$ nuclei shown in Fig.~\ref{fig:nucl:res} and the energy per particle from Fig.~\ref{fig:enerSM} at around $n_\sat$.
This shows the large impact of the saturation properties in uniform matter on the experimental nuclear masses, as expected.

In the following, the different CLDM constructed upon $\chi_{EFT}$ are weighted according to their reproduction of the experimental nuclear masses: H2 and H3 are the best $\chi_{EFT}$ models and define the dark-gray bands plotted for NS crust predictions, then come H1 and H4 which define the light-gray bands, and include the dark-gray one as expected. The others (H5, H7, DHS$_{L59}$ and DHS$_{L69}$) are represented but they will depart from the best-model bands.

\section{Neutron star crust constrained by $\chi_{EFT}$ and experimental nuclear masses}
\label{sec:crust}

We now come to the description of the crust, where the CLDM model for finite nuclei is completed by the contributions of the neutron fluid and of the electrons gas, which also interact with the protons by the Coulomb interaction. At variance with isolated nuclei, nuclear systems in NS crust are at $\beta$-equilibrium and the Wigner-Seitz cell components respect charge neutrality. Details of the model are given in Ref.~\cite{Grams2021}.

The impact of nuclear physics uncertainties on the NS crust has already been analyzed in various ways. Steiner~\cite{Steiner2008}, for instance, has constructed several NS crust using inputs from current experimental information while allowing exploration of the EoS uncertainties, in particular the one induced by the symmetry energy. Tews~\cite{Tews2016} has analyzed the impact of $\chi_{EFT}$ predictions in NM introducing bands reflecting $\chi_{EFT}$ uncertainties. Other approaches have been constructed from currently available EoSs, which may or may not consider the actual predictions for low-density neutron matter~\cite{Baym1971,DouchinHaensel2001,Carreau2019a,Newton2012,Fantina2013,Vinas2017,Carreau2020}.

In the present work, we explore both the theoretical uncertainties in the nuclear matter equation of state from many-body approach based on $\chi_{EFT}$ NN and 3N interactions, as well as the confrontation with the experimental nuclear masses. This is the first approach which investigates in a systematical way these two types of constraints, considering the latest $\chi_{EFT}$ calculations and using the confrontation with experimental nuclear masses to select among the best models for NS crust predictions.

\begin{figure*}[t]
\centering
\includegraphics[scale=0.36]{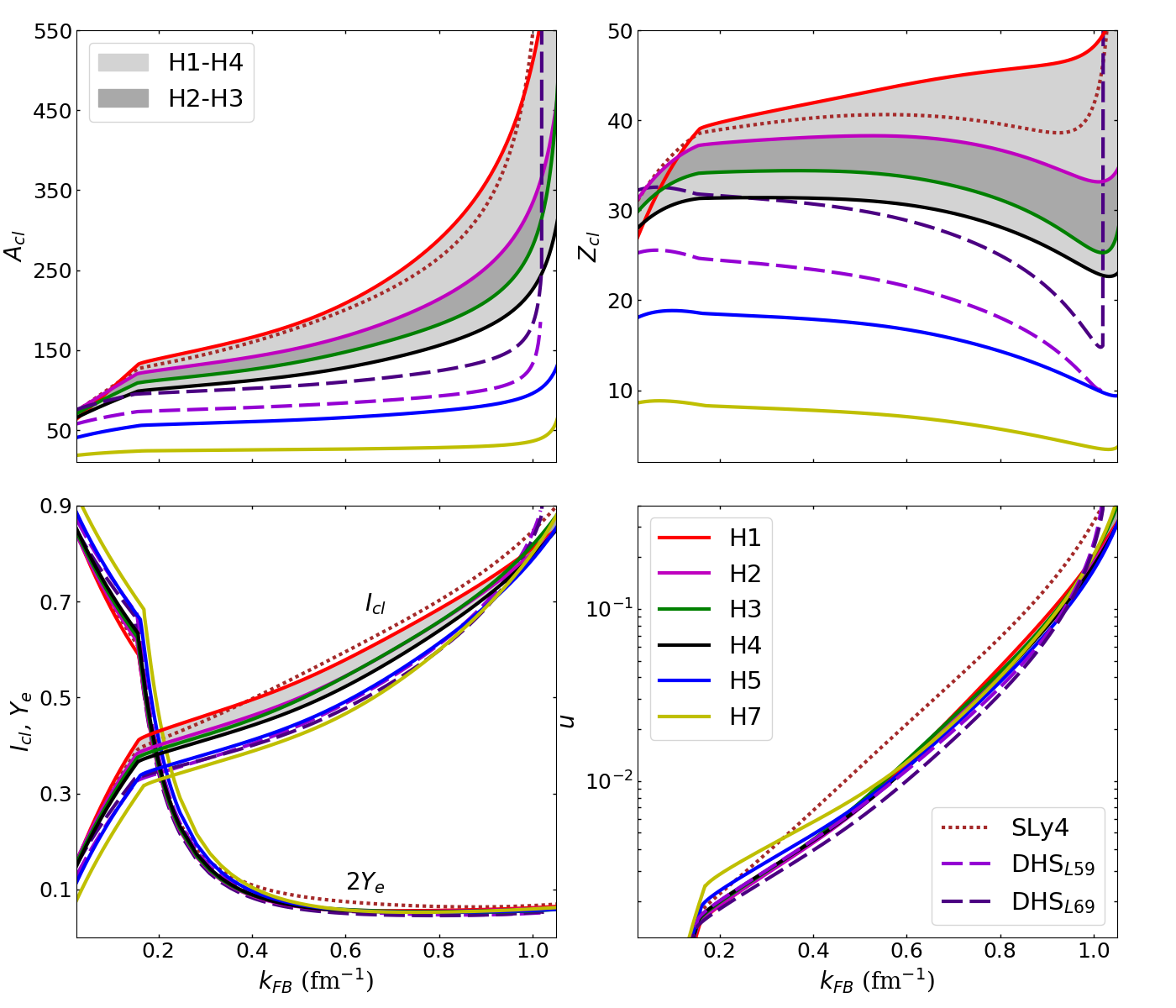}
\caption{Composition (top), asymmetry and $2Y_e$ (bottom left) and volume fraction occupied by nuclei (bottom right) on the neutron star crust as function of the Fermi momentum.}
\label{fig:crust}
\end{figure*}

We display in Fig. \ref{fig:crust} some properties of the NS crust determined by the $\chi_{EFT}$ Hamiltonians and the SLy4 phenomenological force. On the top panels we show the predictions for the composition: the cluster mass number $A_\cl$ (left) and proton number $Z_\cl$ (right) through the crust. Note that $A_\cl$ and $Z_\cl$ are the masses and proton numbers associated to the nuclear cluster only, excluding the contribution of the neutron fluid. Here we prefer for consistency and easy comparison among the results to show all our results in terms of the Fermi momentum $k_{FB}$. On the bottom panels we show the cluster asymmetry $I_\cl=(N_\cl-Z_\cl)/A_\cl$ and electron fraction $Y_e$ (left panel) and the volume fraction $u=V_\cl/V_\WS$ (right panel).

Concerning the predictions for $A_\cl$, $Z_\cl$, $I_\cl$ and $Y_e$, SLy4 is compatible with the predictions of the $\chi_{EFT}$ models (inside the light-gray band). It is remarkable to observe that the predictions of SLy4 are located between the ones of H1 and H2, as already observed in Fig.~\ref{fig:enerSM} for SM. One can then conclude that these predictions are tightly connected to the SM properties of these models, itself closely related to the experimental nuclear masses, and only loosely related to the NM predictions.

The volume fraction $u$ however shows a marked difference between the prediction of SLy4 and the ones from the $\chi_{EFT}$, which are well grouped together. The volume fraction $u$ predicted by SLy4 is larger that the chiral EFT group by about 20-30\%.

\begin{figure}[t]
\centering
\includegraphics[scale=0.25]{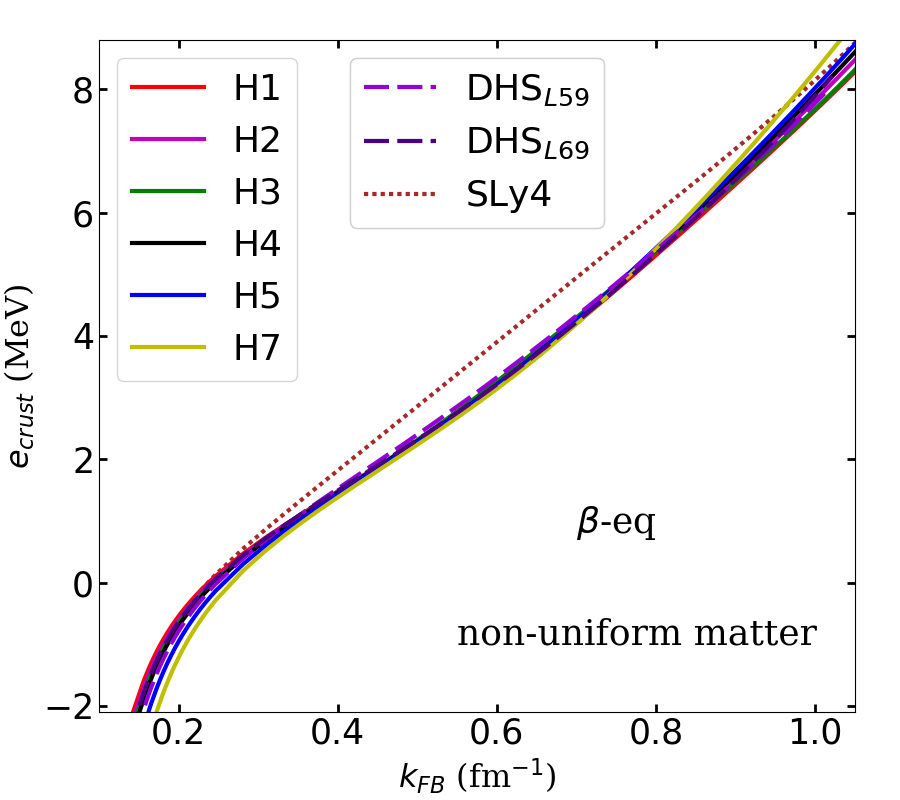}
\includegraphics[scale=0.25]{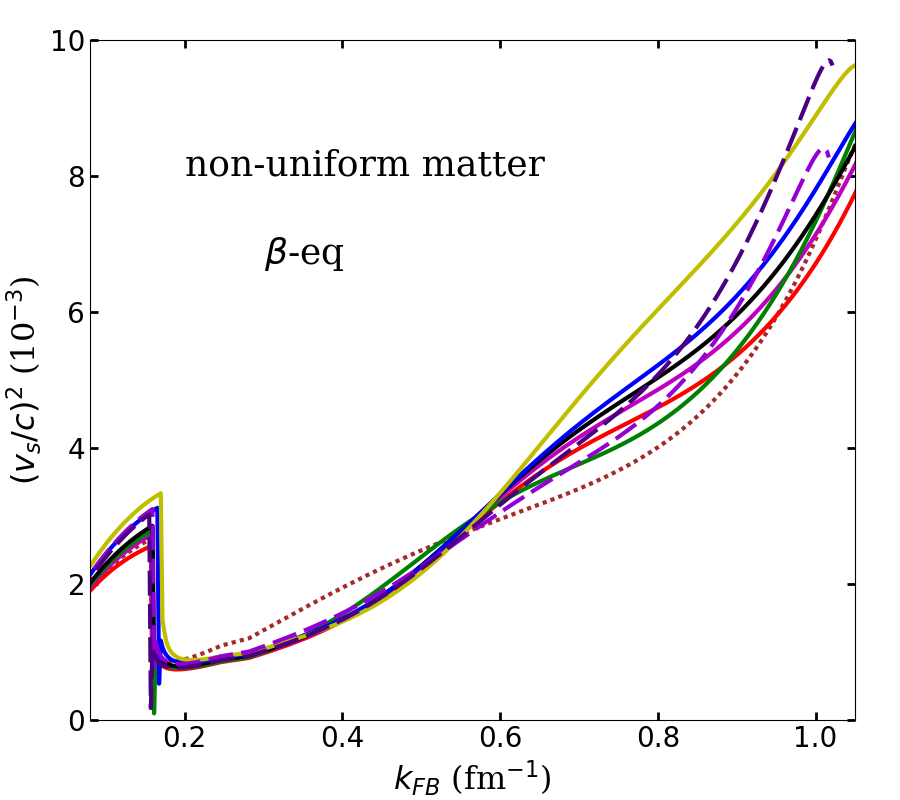}
\caption{Left: energy per particle in the crust for the $\chi_{EFT}$ Hamiltonians and SLy4 phenomenological force. Right: sound speed in the crust predicted by the same models.}
\label{fig:crust:res}
\end{figure}

\begin{figure}[t]
\centering
\includegraphics[scale=0.35]{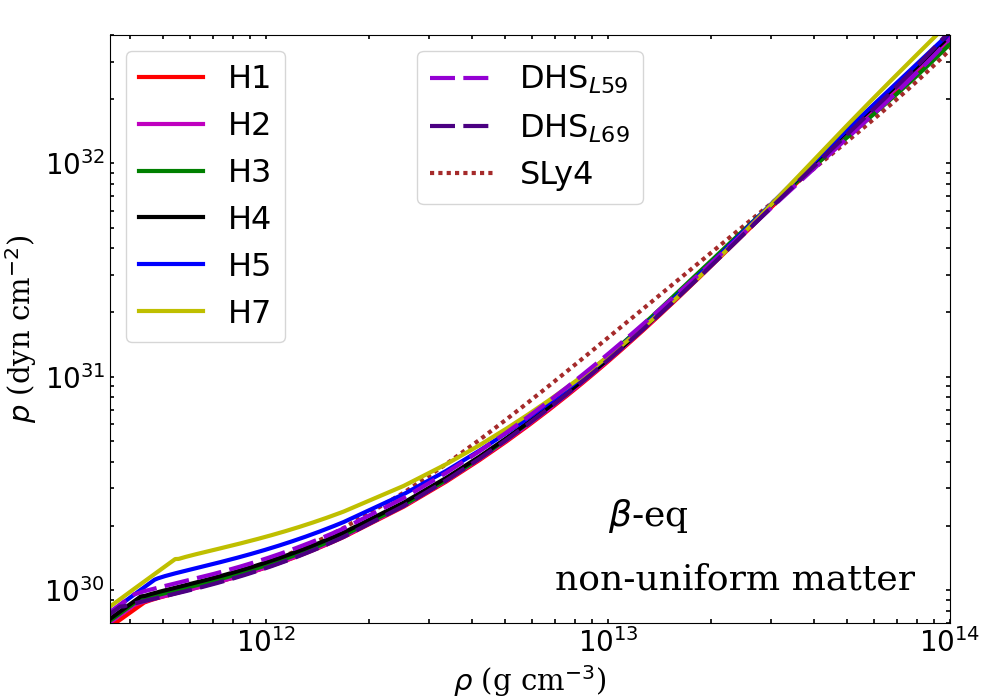}
\caption{EoS (pressure versus energy density) in cgs units for the $\chi_{EFT}$ Hamiltonians and SLy4 phenomenological force.}
\label{fig:crust:eos}
\end{figure}

Other quantities impacted by the NM properties are the energy per particle in the crust, the pressure and the sound speed. The energy per particle and the sound speed are plotted in Fig.~\ref{fig:crust:res} and the EoS, i.e. the pressure as a function of the energy density, in Fig.~\ref{fig:crust:eos}.
We observe that in the density range where $k_F<1.0$~fm$^{-1}$ the prediction for the energy per particle from SLy4 force is about 10-30\% different from the predictions of the group of $\chi_{EFT}$ Hamiltonians, see Fig.~\ref{fig:crust:res}.
The pressure being defined as the density derivative of the energy per particle is also impacted by this marked effect, see Fig.~\ref{fig:crust:eos}.
Finally, the sound speed is obtained as the derivative of the pressure with respect to the energy density, which for the considered densities is almost identical to the density $n_B$. The sound speed predicted by SLy4 is also markedly different from the group of $\chi_{EFT}$ Hamiltonians. Note however that for $k_{FB}>0.6$~fm$^{-1}$ there is a large spreading in the prediction of $\chi_{EFT}$ Hamiltonians which reduces a bit the effect observed for the energy per particle.

In summary, the present study illustrates that different NS crust observables are related to different properties of uniform matter. For instance, the cluster mass $A_\cl$, charge $Z_\cl$ and asymmetry parameter $I_\cl$ are essentially fixed by the SM or close to SM properties where the NM does not play an important role. These properties are controlled by the nuclear masses, and the models can be sorted according to their reproduction of the experimental data.
NM properties impact however other observables, such as the volume fraction $u$, the energy per particle, as well as the internal pressure and the sound speed.
These latter observables are the ones which are greatly influenced by the NM predictions based on chiral EFT, and their accurate prediction can therefore be explained by the relatively well known properties of NM.

\section{Comparison to predictions from other EoS}

\begin{figure*}[t]
\centering
\includegraphics[scale=0.325]{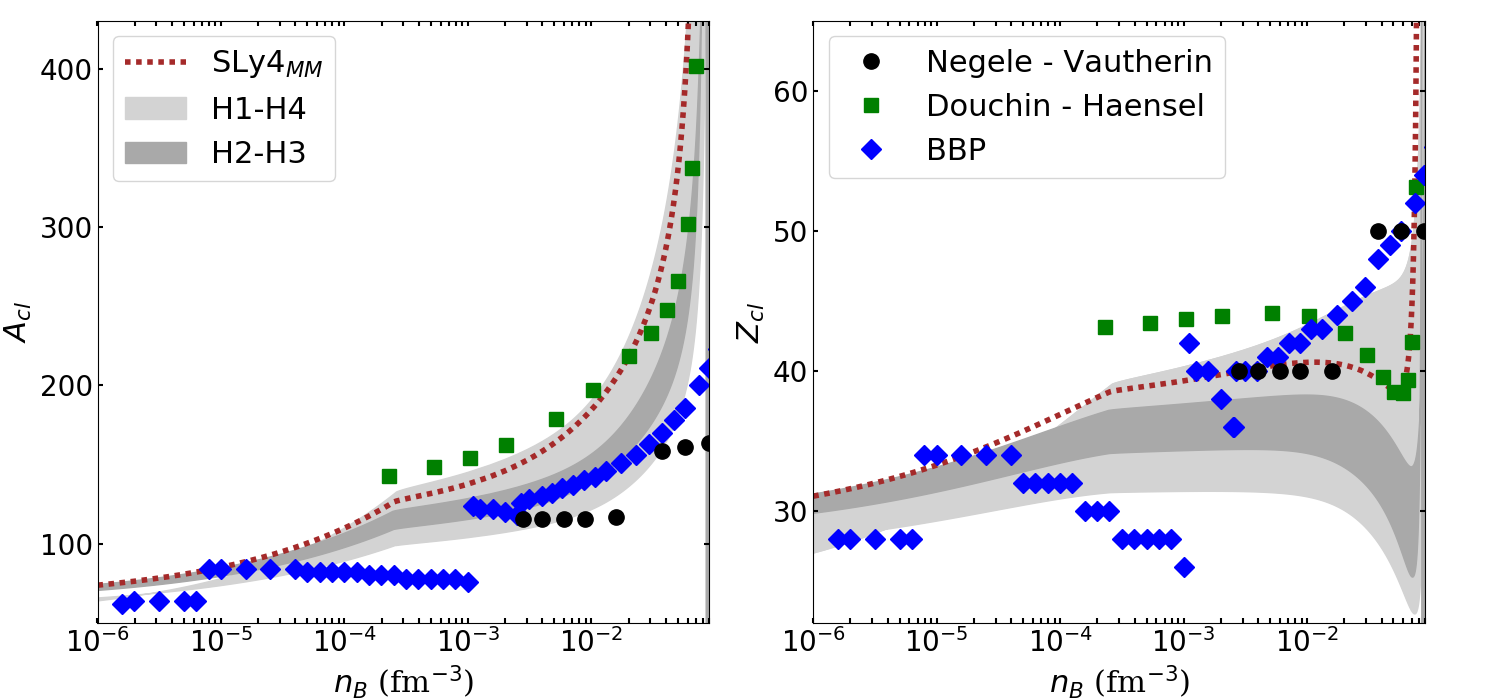}
\caption{Crust composition $A_\cl$, $Z_\cl$ as function of the baryon density. Comparison among  $\chi_{EFT}$ H1-H4, SLy4$_{MM}$ and previous works.}
\label{fig:composition}
\end{figure*}

Finally, we compare our predictions for the NS crust properties to previously determined EoS in Fig.~\ref{fig:composition}.
Our dark-grey contour is given by H2 and H3, while the light-grey contour is defined by H1 and H4. The SLy4 crust properties are shown in dotted red cruve, and can be compared to the Douchin-Haensel predictions in green squares, which are based on the same interaction model.
As a consequence, the SLy4 and Douchin-Haensel EoS are very close, and the small differences which are observed are due to small differences in the CLDM model.

For completeness, we represent in Fig.~\ref{fig:composition} the seminal BBP~\cite{Baym1971} and Negele-Vautherin~\cite{Negele1973} EoS which are the first realistic predictions for the NS crust properties. Our light-grey band is compatible with these other EoS shown in Fig.~\ref{fig:composition}.
The dark-grey band is however narrower and represents the actual uncertainties in the NS crust properties by combining together $\chi_{EFT}$ calculation and the experimental nuclear masses.

\section{Conclusions}

In this study, we have shown that combining together the latest predictions for the uniform matter EoS from $\chi_{EFT}$ approaches with the experimental nuclear masses allows predictions of the NS crust properties which are narrower than the actual uncertainty resulting from the confrontation of various EoS previously known. These predictions were indeed made, for most of them, before the recent achievement from the $\chi_{EFT}$ approach. One can therefore attribute the reduction of the uncertainties in the predictions of the NS crust properties to the advances in the modeling of low-density nuclear matter made possible by $\chi_{EFT}$. 

In addition, the prediction for the NM EoS at low density, where the unitary limit plays an important role in constraining the matter properties, impacts some NS observables, such as the volume fraction $u$, the energy per particle, the pressure and the sound speed.
In the future, it will be interesting to study more closely the role of the unitary limit constrain in NM on these properties.

\begin{acknowledgements}
G.G., J.M. and R.S. are supported by CNRS grant PICS-08294 VIPER (Nuclear Physics for Violent Phenomena in the Universe), the CNRS IEA-303083 BEOS project, the CNRS/IN2P3 NewMAC project, and benefit from PHAROS COST Action MP16214. S.R. is supported by Grant No. DE-FG02-00ER41132 from the Department of Energy , and the Grant No. PHY-1430152 (JINA Center for the Evolution of the Elements), and PHY-1630782 (Network for Neutrinos, Nuclear Astrophysics, and Symmetries (N3AS)) from the National Science Foundation. This work is supported by the LABEX Lyon Institute of Origins (ANR-10-LABX-0066) of the \textsl{Universit\'e de Lyon} for its financial support within the program \textsl{Investissements d'Avenir} (ANR-11-IDEX-0007) of the French government operated by the National Research Agency (ANR).
\end{acknowledgements}


%

\end{document}